\documentclass[letterpaper, 10 pt, conference]{ieeeconf}  % Comment this line out if you need a4paper

\IEEEoverridecommandlockouts                              % This command is only needed if 
\usepackage{tikz}

\usepackage{algorithm}      
\usepackage{algpseudocode}  

\usepackage{booktabs} 

\usepackage{amsmath,amssymb,amsfonts}

\usepackage{siunitx}

\usepackage[acronym]{glossaries}

\newglossaryentry{p_C}{
    name = \ensuremath{p_\mathrm{C}},
    description = {Pressure measured inside the cavity of the mold}
}

\newglossaryentry{c}{
    name = \ensuremath{c},
    description = {Cycle number}
}

\newglossaryentry{p_S}{
    name = \ensuremath{p_\mathrm{S}},
    description = {Pressure measured inside the screw antechamber}
}

\newglossaryentry{Q_S}{
    name = \ensuremath{Q_\mathrm{S}},
    description = {Volumetric flow rate induced by the axial screw velocity}
}

\newglossaryentry{V_S}{
    name = \ensuremath{V_\mathrm{S}},
    description = {Volume of the screw antechamber}
}

\newglossaryentry{V_D}{
    name = \ensuremath{V_\mathrm{D}},
    description = {dosing volume for the screw antechamber}
}

\newglossaryentry{phi}{
    name = \ensuremath{\varphi},
    description = {Degree of filling of the cavity}
}

\newglossaryentry{V_C}{
    name = \ensuremath{V_\mathrm{C}},
    description = {Volume of the cavity}
}

\newglossaryentry{u_S}{
    name = \ensuremath{u_\mathrm{S}},
    description = {Screw voltage}
}

\newglossaryentry{u_ANN}{
    name = \ensuremath{\boldsymbol{u}_\mathrm{ANN}},
    description = {inputs to the artificial neural network}
}

\newglossaryentry{y_ANN}{
    name = \ensuremath{\boldsymbol{y}_\mathrm{ANN}},
    description = {outputs of the artificial neural network}
}

\newglossaryentry{A_D}{
    name = \ensuremath{\mathbf{A}_{\mathrm{D}}},
    description = {time-discrete system matrix}
}

\newglossaryentry{B_D}{
    name = \ensuremath{\boldsymbol{B}_{\mathrm{D}}},
    description = {time-discrete input matrix}
}

\newglossaryentry{E_D}{
    name = \ensuremath{\boldsymbol{E}_{\mathrm{D}}},
    description = {time-discrete matrix for constant terms}
}

\newglossaryentry{P}{
    name = \ensuremath{\boldsymbol{P}},
    description = {constant term inside the prediction equation}
}

\newglossaryentry{Phi}{
    name = \ensuremath{\mathbf{\Phi}},
    description = {transition matrix}
}

\newglossaryentry{Kappa}{
    name = \ensuremath{\pmb{\kappa}},
    description = {kappa matrix for the calculation of the state vector}
}

\newglossaryentry{Zeta}{
    name = \ensuremath{\pmb{\zeta}},
    description = {zeta matrix for the calculation of the cavity-pressure vector from the state vector}
}

\newglossaryentry{p_C_ref}{
    name = \ensuremath{p_\mathrm{C}^{\mathrm{ref}}},
    description = {Cavity-pressure reference}
}

\newglossaryentry{p_C_ref_min}{
    name = \ensuremath{p_\mathrm{C}^{\mathrm{ref, min}}},
    description = {Minimum cavity-pressure reference}
}

\newglossaryentry{p_C_ref_max}{
    name = \ensuremath{p_\mathrm{C}^{\mathrm{ref, max}}},
    description = {Maximum cavity-pressure reference}
}

\newglossaryentry{K_D}{
    name = \ensuremath{K_\mathrm{D}},
    description = {Gain of the drive model}
}

\newglossaryentry{T_D}{
    name = \ensuremath{T_\mathrm{D}},
    description = {Time constant of the drive model}
}

\newglossaryentry{C_1}{
    name = \ensuremath{C_1},
    description = {first coefficient for the equation of the volumetric flow rate}
}

\newglossaryentry{C_2}{
    name = \ensuremath{C_2},
    description = {Power law index for the equation of the volumetric flow rate}
}

\newglossaryentry{C_3}{
    name = \ensuremath{C_3},
    description = {Third coefficient for the equation of the volumetric flow rate}
}

\newglossaryentry{Q_C}{
    name = \ensuremath{Q_\mathrm{C}},
    description = {Volumetric flow rate entering the cavity of the mold}
}

\newglossaryentry{Q_L}{
    name = \ensuremath{Q_\mathrm{L}},
    description = {leakage volumetric flow rate from the screw antechamber towards the feed zone of the screw}
}

\newglossaryentry{Delta_V_S}{
    name = \ensuremath{\Delta V_\mathrm{S}},
    description = {Displaced volume of the screw}
}

\newglossaryentry{Delta_V_R}{
    name = \ensuremath{\Delta V_\mathrm{R}},
    description = {Displaced volume of the screw where the cavity pressure rise is expected}
}

\newglossaryentry{K_S}{
    name = \ensuremath{ K_\mathrm{S}},
    description = {Bulk modulus inside the screw antechamber}
}

\newglossaryentry{K_C}{
    name = \ensuremath{ K_\mathrm{C}},
    description = {Bulk modulus inside the cavity of the mold}
}

\newglossaryentry{x}{
    name = \ensuremath{ \boldsymbol{x}},
    description = {state vector}
}

\newglossaryentry{x_inter}{
    name = \ensuremath{ \boldsymbol{x}_{\mathrm{inter}}},
    description = {intermediate state vector}
}

\newglossaryentry{N_x}{
    name = \ensuremath{N_{\boldsymbol{x}}},
    description = {number of state variables}
}

\newglossaryentry{N_y}{
    name = \ensuremath{N_{\boldsymbol{y}}},
    description = {number of output variables}
}

\newglossaryentry{N_d}{
    name = \ensuremath{N_\mathrm{d}},
    description = {number of series elements for calculation of the time-discrete system matrices}
}

\newglossaryentry{N_Theta}{
    name = \ensuremath{N_{\boldsymbol{\Theta}}},
    description = {number of model optimization parameters}
}

\newglossaryentry{N_xi}{
    name = \ensuremath{N_{\boldsymbol{\xi}}},
    description = {number of uncondensed optimization parameters}
}

\newglossaryentry{S_dot_S}{
    name = \ensuremath{\dot{S}_\mathrm{S}},
    description = {source term for the screw-pressure-build-up equation}
}

\newglossaryentry{S_dot_C}{
    name = \ensuremath{\dot{S}_\mathrm{C}},
    description = {source term for the cavity-pressure-build-up equation}
}

\newglossaryentry{t}{
    name = \ensuremath{t},
    description = {time variable}
}

\newglossaryentry{f}{
    name = \ensuremath{\boldsymbol{f}},
    description = {system of ordinary differential equations}
}

\newglossaryentry{x_hat}{
    name = \ensuremath{ \hat{\boldsymbol{x}}},
    description = {estimated state vector}
}

\newglossaryentry{x_O}{
    name = \ensuremath{ {\boldsymbol{x}}_\mathrm{O}},
    description = {operating state for linearization}
}

\newglossaryentry{u_O}{
    name = \ensuremath{ {u}_\mathrm{O}},
    description = {operating input for linearization}
}

\newglossaryentry{u}{
    name = \ensuremath{u},
    description = {input}
}

\newglossaryentry{y}{
    name = \ensuremath{ \boldsymbol{y}},
    description = {output vector}
}

\newglossaryentry{N_psi}{
    name = \ensuremath{ {N_\psi}},
    description = {number of basis functions}
}

\newglossaryentry{psi_i}{
    name = \ensuremath{ {\psi_i}},
    description = {basis function}
}

\newglossaryentry{psi}{
    name = \ensuremath{ \boldsymbol{\psi}},
    description = {vector of basis functions}
}

\newglossaryentry{w_i}{
    name = \ensuremath{ {w_i}},
    description = {specific weight for the basis function}
}

\newglossaryentry{w}{
    name = \ensuremath{\boldsymbol{w}},
    description = {weight vector for basis functions}
}

\newglossaryentry{r_i}{
    name = \ensuremath{r_i},
    description = {center for basis function}
}

\newglossaryentry{a}{
    name = \ensuremath{a},
    description = {variable to set the steepness of the basis function}
}

\newglossaryentry{Theta}{
    name = \ensuremath{\boldsymbol{\Theta}},
    description = {Parameter vector of the model}
}

\newglossaryentry{Theta_C}{
    name = \ensuremath{\boldsymbol{\theta}_\mathrm{C}},
    description = {Controller Parameters}
}

\newglossaryentry{m}{
    name = \ensuremath{m},
    description = {part mass}
}

\newglossaryentry{m_ref}{
    name = \ensuremath{m^{\mathrm{ref}}},
    description = {reference for the part mass}
}

\newglossaryentry{Delta_t}{
    name = \ensuremath{\Delta t},
    description = {time step size}
}

\newglossaryentry{N_1}{
    name = \ensuremath{N_1},
    description = {Length of the prediction horizon inside the MPC}
}

\newglossaryentry{N_2}{
    name = \ensuremath{N_2},
    description = {Length of the control horizon}
}

\newglossaryentry{N_C}{
    name = \ensuremath{N_\mathrm{C}},
    description = {Length of the cycle (time steps)}
}

\newglossaryentry{X}{
    name = \ensuremath{\boldsymbol{X}},
    description = {Complete state vector (containing multiple time steps)}
}

\newglossaryentry{p_C_vec}{
    name = \ensuremath{\boldsymbol{p}_\mathrm{C}},
    description = {Complete cavity-pressure vector (containing multiple time steps)}
}

\newglossaryentry{p_C_ref_vec}{
    name = \ensuremath{\boldsymbol{p}_\mathrm{C}^\mathrm{ref}},
    description = {Complete cavity-pressure reference vector (containing multiple time steps)}
}

\newglossaryentry{p_C_ref_star}{
    name = \ensuremath{p_\mathrm{C}^{\mathrm{ref, *}}},
    description = {Optimal cavity-pressure reference}
}

\newglossaryentry{U}{
    name = \ensuremath{\boldsymbol{U}},
    description = {Complete input vector (containing multiple time steps)}
}

\newglossaryentry{xi}{
    name = \ensuremath{\boldsymbol{\xi}},
    description = {Complete vector of optimization variables (containing multiple time steps)}
}

\newglossaryentry{n}{
    name = \ensuremath{n},
    description = {Optimization iteration}
}

\newglossaryentry{eta}{
    name = \ensuremath{{\eta}},
    description = {Slack variable}
}

\newglossaryentry{alpha_SQP}{
    name = \ensuremath{{\alpha}_\mathrm{SQP}},
    description = {step size of the SQP}
}

\newglossaryentry{Q_mat}{
    name = \ensuremath{\mathbf{Q}},
    description = {Matrix for tracking penalty}
}

\newglossaryentry{Q}{
    name = \ensuremath{Q},
    description = {Value for tracking penalty}
}

\newglossaryentry{R_mat}{
    name = \ensuremath{\mathbf{R}},
    description = {Matrix for actuator penalty}
}

\newglossaryentry{R}{
    name = \ensuremath{R},
    description = {Value for actuator penalty}
}

\newglossaryentry{rho}{
    name = \ensuremath{\rho},
    description = {weight for the slack variable}
}

\newglossaryentry{b_eq_MPC}{
    name = \ensuremath{\boldsymbol{b}_\mathrm{eq}^{\mathrm{MPC}}},
    description = {vector of equality constraints for the model predictive controller}
}

\newglossaryentry{b_ieq_MPC}{
    name = \ensuremath{\boldsymbol{b}_\mathrm{ieq}^{\mathrm{MPC}}},
    description = {vector of inequality constraints for the model predictive controller}
}

\newglossaryentry{i}{
    name = \ensuremath{i},
    description = {step variable inside the prediction horizon}
}

\newglossaryentry{u_S_upperBound}{
    name = \ensuremath{u_\mathrm{S, lim}^\mathrm{+}},
    description = {upper bound of the screw voltage}
}

\newglossaryentry{u_S_lowerBound}{
    name = \ensuremath{u_\mathrm{S, lim}^\mathrm{-}},
    description = {lower bound of the screw voltage}
}

\newglossaryentry{p_C_lim_vec}{
    name = \ensuremath{\boldsymbol{p}_\mathrm{C, lim}},
    description = {upper bound for cavity pressure}
}

\newglossaryentry{p_C_R}{
    name = \ensuremath{p_{\mathrm{C}}^{\mathrm{R}}},
    description = {pressure where mode is switched to packing}
}

\newglossaryentry{r_MPC}{
    name = \ensuremath{\boldsymbol{r}_\mathrm{MPC}},
    description = {residual for the Gauss-Newton method for the model predictive controller}
}

\newglossaryentry{H_QP_MPC}{
    name = \ensuremath{\mathbf{H}_\mathrm{QP}^{\mathrm{MPC}}},
    description = {Approximated uncondensed Hessian of the sequential quadratic programming approach}
}

\newglossaryentry{f_QP_MPC}{
    name = \ensuremath{\boldsymbol{f}_\mathrm{QP}^{\mathrm{MPC}}},
    description = {uncondensed gradient of the sequential quadratic programming approach}
}

\newglossaryentry{Delta_xi}{
    name = \ensuremath{\Delta \boldsymbol{\xi}},
    description = {deviation of the optimization variables from the current best guess of \gls*{xi}}
}

\newglossaryentry{A_eq_MPC}{
    name = \ensuremath{\mathbf{A}_\mathrm{eq}^{\mathrm{MPC}}},
    description = {Uncondensed linear equality matrix}
}

\newglossaryentry{A_ieq_MPC}{
    name = \ensuremath{\mathbf{A}_\mathrm{ieq}^{\mathrm{MPC}}},
    description = {Uncondensed linear inequality matrix}
}

\newglossaryentry{Delta_X}{
    name = \ensuremath{\Delta \boldsymbol{X}},
    description = {deviation of the optimized states from the current best guess of \gls*{X}}
}

\newglossaryentry{Delta_U}{
    name = \ensuremath{\Delta \boldsymbol{U}},
    description = {deviation of the optimized inputs from the current best guess of \gls*{U}}
}

\newglossaryentry{Delta_eta}{
    name = \ensuremath{\Delta \eta},
    description = {deviation of the optimized slack from the current best guess of \gls*{eta}}
}

\newglossaryentry{delta_U}{
    name = \ensuremath{\delta \boldsymbol{U}},
    description = {deviation of the optimized inputs inside the control horizon}
}

\newglossaryentry{kappa}{
    name = \ensuremath{\kappa},
    description = {rediscretization factor}
}

\newglossaryentry{T_int}{
    name = \ensuremath{T_\mathrm{int}},
    description = {integration time}
}

\newglossaryentry{1}{
    name = \ensuremath{\boldsymbol{1}},
    description = {identity vector}
}

\newglossaryentry{Delta_p_C_OS}{
    name = \ensuremath{\Delta p_\mathrm{C, OS}},
    description = {maximum allowed overshoot of the cavity-pressure reference}
}

\newglossaryentry{K_P}{
    name = \ensuremath{K_\mathrm{P}},
    description = {Gain of the proportional controller}
}

\newglossaryentry{N_SQP_MPC}{
    name = \ensuremath{N_\mathrm{SQP}^{\mathrm{MPC}}},
    description = {Number of SQP steps inside the model predictive controller}
}

\newglossaryentry{r_MO}{
    name = \ensuremath{\boldsymbol{r}_\mathrm{MO}},
    description = {residual for the model optimizer}
}

\newglossaryentry{N_B}{
    name = \ensuremath{N_\mathrm{B}},
    description = {length of the recording}
}

\newglossaryentry{N_SQP_MO}{
    name = \ensuremath{N_\mathrm{SQP}^{\mathrm{MO}}},
    description = {Number of SQP steps inside the model optimizer}
}

\newglossaryentry{A_ieq_MO}{
    name = \ensuremath{\mathbf{A}_\mathrm{ieq}^{\mathrm{MO}}},
    description = {linear inequality matrix for the model optimizer}
}

\newglossaryentry{b_ieq_MO}{
    name = \ensuremath{\boldsymbol{b}_\mathrm{ieq}^{\mathrm{MO}}},
    description = {inequality constraints for the model optimizer}
}

\newglossaryentry{H_QP_MO}{
    name = \ensuremath{\mathbf{H}_\mathrm{QP}^{\mathrm{MO}}},
    description = {Hessian Matrix for the model optimization}
}

\newglossaryentry{f_QP_MO}{
    name = \ensuremath{\boldsymbol{f}_\mathrm{QP}^{\mathrm{MO}}},
    description = {gradient vector for the model optimization}
}

\newglossaryentry{sigmaW}{
    name = \ensuremath{\sigma_\mathrm{W}},
    description = {forgetting factor for the time-variant Bayesian Optimization}
}

\newglossaryentry{sigma0}{
    name = \ensuremath{\sigma_{0}},
    description = {}
}

\newglossaryentry{ell}{
    name = \ensuremath{l},
    description = {}
}

\newglossaryentry{k}{
    name = \ensuremath{k},
    description = {kernel or covariance function of the gaussian process}
}

\newglossaryentry{k_SE}{
    name = \ensuremath{k_\mathrm{SE}},
    description = {squared exponential kernel or covariance function}
}

\newglossaryentry{k_W}{
    name = \ensuremath{k_\mathrm{W}},
    description = {Wiener kernel or covariance function}
}

\newglossaryentry{sigma}{
    name = \ensuremath{\sigma},
    description = {standard deviation of the measurements}
}

\newglossaryentry{Theta_GPR}{
    name = \ensuremath{\boldsymbol{\Theta}_\mathrm{GPR}},
    description = {hyperparameters of the Gaussian process regression model}
}

\newglossaryentry{alpha}{
    name = \ensuremath{\alpha},
    description = {acquisition function for the Bayesian optimization}
}

\newglossaryentry{eps}{
    name = \ensuremath{\epsilon},
    description = {acquisition function for the Bayesian optimization}
}

\newglossaryentry{kappa_suc}{
    name = \ensuremath{\kappa_\mathrm{suc}},
    description = {Success parameter for the local BO}
}

\newglossaryentry{kappa_fail}{
    name = \ensuremath{\kappa_\mathrm{fail}},
    description = {Failure parameter for the local BO}
}

\newglossaryentry{GP_corr}{
    name = \ensuremath{\mathcal{GP}_{\mathrm{corr}}},
    description = {correction model for the costs}
}

\newglossaryentry{f_P}{
    name = \ensuremath{\boldsymbol{f}_\mathrm{P}},
    description = {state space model for the pressure}
}

\newglossaryentry{f_D}{
    name = \ensuremath{\boldsymbol{f}_\mathrm{D}},
    description = {state space model for the drive}
}

\newglossaryentry{r_D}{
    name = \ensuremath{\boldsymbol{r}_\mathrm{D}},
    description = {residual for the drive model}
}

\newglossaryentry{R_D}{
    name = \ensuremath{\boldsymbol{R}_\mathrm{D}},
    description = {total residual for the drive model}
}

\newglossaryentry{r_P}{
    name = \ensuremath{\boldsymbol{r}_\mathrm{P}},
    description = {residual for the pressure model}
}

\newglossaryentry{R_P}{
    name = \ensuremath{\boldsymbol{R}_\mathrm{P}},
    description = {total residual for the pressure model}
}

\newglossaryentry{N_T}{
    name = \ensuremath{N_\mathrm{T}},
    description = {Number of training data points}
}

\newglossaryentry{w_D}{
    name = \ensuremath{\boldsymbol{w}_\mathrm{D}},
    description = {weights and biases of the ANN for the Drive}
}

\newglossaryentry{w_1}{
    name = \ensuremath{\boldsymbol{w}_1},
    description = {weights and biases of the ANN for the first pressure model}
}

\newglossaryentry{w_2}{
    name = \ensuremath{\boldsymbol{w}_2},
    description = {weights and biases of the ANN for the second pressure model}
}

\newglossaryentry{w_3}{
    name = \ensuremath{\boldsymbol{w}_3},
    description = {weights and biases of the ANN for the third pressure model}
}

\newglossaryentry{lambda_Reg}{
    name = \ensuremath{{\lambda}_\mathrm{Reg}},
    description = {Regularization constant for the time derivative}
}

\newglossaryentry{Theta_C_cand}{
    name = {\ensuremath{\hat{\boldsymbol{\theta}}_\mathrm{C}}}, 
    description = {candidate controller parameters for the next cycle}
}

\newglossaryentry{Theta_C_star}{
    name = \ensuremath{\boldsymbol{\theta}_\mathrm{C}^{\mathrm{*}}},
    description = {Best controller parameters so far}
}

\newglossaryentry{J_M}{
    name = \ensuremath{J_\mathrm{M}},
    description = {Cost function defined by the simulation of the control loop}
}

\newglossaryentry{n_LM}{
    name = \ensuremath{n_\mathrm{LM}},
    description = {Number of Levenberg-Marquardt steps for the model fitting}
}

\newglossaryentry{n_C}{
    name = \ensuremath{n_\mathrm{C}},
    description = {Number of cycles used for model fitting}
}

\newglossaryentry{b}{
    name = \ensuremath{\boldsymbol{b}},
    description = {Box constraint}
}

\newcommand{\RNum}[1]{\uppercase\expandafter{\romannumeral #1\relax}}

\newcommand{\ShowFontSize}{\f@size pt}

\DeclareSIUnit\bar{bar}

\def\BibTeX{{\rm B\kern-.05em{\sc i\kern-.025em b}\kern-.08em
    T\kern-.1667em\lower.7ex\hbox{E}\kern-.125emX}}

\title{\LARGE \bf
Model-Guided Local Bayesian Optimization for Tuning of Interpretable Controllers in Injection Molding
}

\author{Jens Ahlers$^{1}$, Robert Göllinger$^{1}$, Xu Chen$^{1}$, Heike Vallery$^{1,2}$, and Sebastian Stemmler$^{1}$% <-this % stops a space
\thanks{*The presented research was funded by the Deutsche Forschungsgemeinschaft (German Research Foundation) under the funding code 378417139 (Phasenübergreifende Prozessführungskonzepte beim Spritzgießen unter Nutzung moderner Regelungsstrategien).}% <-this % stops a space
\thanks{
    $^{1}$Institute of Automatic Control, RWTH Aachen University, Aachen, Germany; \newline
    \indent$^{2}$Faculty of Mechanical Engineering, TU Delft, Delft, The Netherlands
    {\tt\small \{j.ahlers, r.goellinger, x.chen, h.vallery, s.stemmler\}@irt.rwth-aachen.de}}%    
}
\begin{document}

\maketitle
\thispagestyle{empty}
\pagestyle{empty}

%%%%%%%%%%%%%%%%%%%%%%%%%%%%%%%%%%%%%%%%%%%%%%%%%%%%%%%%%%%%%%%%%%%%%%%%%%%%%%%%
\begin{abstract}
% Introduction: 
Advanced control methods have proven effective for controlling cavity pressure, a key determinant of part-quality attributes, in the plastics injection molding process.
However, the abstract nature of the resulting control laws makes them difficult to interpret in a production environment, thereby limiting adoption in industrial applications.
Additionally, controller optimization poses a severe challenge due to the diversity of mold geometries and materials.

% Methods: 2 Sentences
We propose a method to automatically optimize interpretable controllers during manufacturing while being cycle-efficient and risk-aware.
The approach uses a Physics-Inspired Neural Mixture-of-Local-Experts model of the injection molding dynamics and augments its simulated closed-loop costs with a residual Gaussian Process, enabling Local Bayesian Optimization of controller parameters.

% Results and Discussion: 2 Sentences
We benchmark the algorithm against Vanilla Bayesian Optimization (BO) in simulation, using three controllers with parameter counts ranging from 1 to 30.
Using the local method, we identify controller parameters that yield costs comparable to or lower than those of global BO over 20 optimization iterations, while mitigating high-cost excursions during tuning.

% Classified that it is indeed a local method. But for our tests, we found the same optimal solution
% Tried not to blame the machine operators 
% Added some specifics (how many parameters and so on)
\end{abstract}

\section{Introduction} \label{sec:introduction}

% relevance of injection molding: 
Injection molding (IM) is one of the most vital processes in the plastics processing industry, enabling the large-scale production of almost arbitrarily complex plastic products~\cite{hopmannPlasticsIndustry402021}.
With worldwide annual production exceeding 400 million tons and continuing to rise, the efficiency and quality of this process are critical~\cite{geyerProductionUseFate2020}.
 
% Common control concepts
Usually in industrial IM, control schemes switch between screw-velocity control during the injection phase and screw-pressure control during the packing phase~\cite{hornbergSwitchoverChallengeIts2020}.
Determining the switch-over point between the two control strategies remains a topic of current research~\cite{bielenbergManualAutomatedExploring2025}. 

% Importance of cavity pressure:  
Although in-mold cavity-pressure sensors are not considered an industrial standard in IM, the relationships between cavity pressure and part-quality attributes have been widely studied in the literature~\cite{kurtExperimentalInvestigationPlastic2009}. 
Due to strong correlations between the cavity pressure and part-quality attributes, cavity pressure is frequently used as a feature in quality-prediction models~\cite{locknerTransferLearningArtificial2022}.
Some approaches even treat cavity pressure as the controlled variable~\cite{stemmlerQualityControlInjection2020}.

% Controlling the cavity pressure: 
Treating cavity pressure as the controlled variable during the injection and packing phases circumvents the need to determine a switch-over point.
The approaches reported in the literature predominantly rely on dynamic models to perform online or in-between-cycle optimizations to determine a suitable controller output~\cite{vukovicAdaptiveModelbasedPredictive2022}. 

% Hurdles: 
% Ich versuche hier klarzustellen, dass in dem einen Fall sich die Stellgröße direkt aus dem Modell ergibt und in unserem Fall aus einer einfach zu verstehenden Kurve, die allerdings durch BO und Modell parametriert wird.
However, model-based controllers face two challenges to industrial adoption:
\begin{enumerate}
    \item Sophisticated model-based control algorithms almost behave like a black box. 
    This makes it impossible for machine operators to assess and verify controller outputs, leading them to mistrust such controllers~\cite{forbesModelPredictiveControl2015}.
    \item Obtaining accurate and reliable process models is difficult since cavity-pressure dynamics are highly sensitive to mold geometry, sensor location, and material properties~\cite{froehlichControlorientedModelingServopump2018}. 
    Due to the individuality and almost arbitrary complexity of mold geometries, first-principles modeling is cumbersome and can often yield inaccurate models.
    Conversely, relying solely on data-driven models for model-based control can be risky, particularly during the initialization phase when sufficient data are not available.
\end{enumerate}

% Sketching one approach: Interpretable control laws: 
% Written more about interpretable control laws and cited an application paper.
% written that it is a parametrized curve
One approach to address these challenges is to employ interpretable control laws that are parametrized by an optimization algorithm.
In \cite{heinInterpretableControlReinforcement2020}, the authors train human-interpretable control laws, implemented as fuzzy controllers, using reinforcement learning on a cart-pole example.
Interpretable control laws offer a transparent behavior, often visualized as a single parametrized curve, enabling machine operators to quickly assess and verify control decisions.
To compete with model-based control schemes and adapt to different IMPs, the control law must be sufficiently flexible.
To support plug-and-play capability, the algorithm should automatically determine its optimal parameters during manufacturing.
Conducting trials that produce scrap parts is detrimental to economic goals, so this optimization should converge as quickly as possible.

% One approach to optimize them efficiently: 
Bayesian Optimization (BO) is known for data-efficiently optimizing black-box problems by learning a surrogate model of the cost function during optimization~\cite{bishopPatternRecognitionMachine2006}.
While recent work has applied Vanilla BO to the tuning of an adaptive model predictive cavity-pressure controller~\cite{vukovicCrossphaseInjectionMolding2024}, Vanilla BO scales poorly in high-dimensional input spaces~\cite{bishopPatternRecognitionMachine2006}. 

% High-dimensional controller?  
In the literature, multiple methods have been proposed to address the challenges of high-dimensional black-box optimization. 
The TurBO algorithm~\cite{erikssonScalableGlobalOptimization2019}, an
extension to Vanilla BO, focuses on local trust regions rather than maintaining a global surrogate model across the entire search space.
Other extensions to Vanilla BO leverage first-order~\cite{mullerLocalPolicySearch2021} and second-order~\cite{brunzemaBayeSQPBayesianOptimization2026} optimization techniques to determine the next sample point.

% Incorporating domain knowledge:
While these approaches assume the process is a black box, additional information can be incorporated by, e.g., using a composite objective function to guide the tuning process as presented in~\cite{astudilloThinkingBoxTutorial2022}.

% Outlining the approach for IM: 
Effective automated controller tuning for IM requires minimizing the risk of machine damage and scrap-part production while remaining data-efficient. 
To achieve this, we adapt the TurBO-1 algorithm to the Injection Molding Process (IMP), yielding a Model-Guided Local Bayesian Optimization (MGLBO) approach. 
Central to this method is a composite objective function that approximates closed-loop costs by combining two models: 
\begin{enumerate}
    \item a Physics-Inspired Neural Mixture-of-Local-Experts model that approximates the underlying process dynamics, and
    \item a Gaussian Process (GP) regression model that corrects the mismatch between the dynamic model's simulated closed-loop costs and real-world plant observations.
\end{enumerate}
The algorithm computes new controller parameters during the cooling phase of the IMP, utilizing an acquisition function to balance predicted performance against uncertainty.
We investigate this approach for three interpretable controllers: 
\begin{enumerate}
    \item Proportional (P) controller, 
    \item Gain-Scheduled Proportional Integral (G-PI) controller, and
    \item Radial-Basis-Function (RBF) controller.
\end{enumerate} 
In simulations, we benchmark the performance of the MGLBO against Vanilla BO. 

% Clearly stating the contribution: 
To our knowledge, this is the first work to propose an IM process-controller-optimization approach that leverages a dynamic process model and plant observations to achieve both data efficiency and risk awareness.

% Outline the structure of the paper: 
The contribution is structured as follows.
First, the dynamic model is introduced in Sec.~\ref {sec:process-modeling}, which is then integrated into the trust-region optimization using Local BO in Sec.~\ref {sec: TrustRegionOptimization}.
In Sec.~\ref {sec:simulated-experiments}, we define the tested controllers and the simulation setup to assess the effectiveness of MGLBO. Results are presented in Sec.~\ref {sec:results} and discussed in Sec.~\ref{sec:discussion}.
\section{Process Modeling} \label{sec:process-modeling}

\subsection{Overview}

\begin{figure}[ht]
    \centering
    \includegraphics[width=\columnwidth]{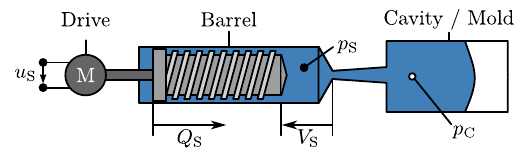}
    \caption{Schematic overview of the IMP.}
    \label{fig:plant}
\end{figure}

% Improved readability 
% thanks for the help
A schematic overview of the plant is presented in Fig.~\ref{fig:plant}. 
The command signal~\gls{u_S} for the frequency inverter of the servo-electric drive of the IM machine serves as control input.
The resulting screw velocity determines the volumetric flow rate~\gls{Q_S}. 
The screw antechamber has volume~\gls{V_S} and 
measured internal pressure~\gls{p_S}, while the cavity pressure~\gls{p_C} inside the mold serves as the controlled variable with its reference~\gls{p_C_ref}.
Resistance in the nozzle and runner causes a pressure drop~\mbox{$\Delta p = \gls{p_S} - \gls{p_C}$} between the screw antechamber and the cavity-pressure sensor.
We choose the system-state vector as~\mbox{$\gls{x} := \begin{bmatrix}\gls{V_S} & \gls{Q_S} & \gls{p_S} & \gls{p_C} \end{bmatrix}^\top$}.
The complete model consists of a drive model and a pressure model.

\subsection{Drive Model}
We approximate the IM's drive dynamics using a Residual Network (ResNet), a common Artificial Neural Network (ANN) architecture for modeling dynamic systems, inspired by explicit Euler integration~\cite{chenNeuralOrdinaryDifferential2018}.
The time-discrete nonlinear state-space equations
% the equation for the drives in vector format: 
\begin{multline}
    \begin{bmatrix}
        \gls*{V_S} (k+1) \\
        \gls*{Q_S} (k+1)
    \end{bmatrix}   = 
    \gls{f_D}(\gls{V_S}(k), \gls{u_S}(k), \gls{Q_S}(k)) = \\
    \begin{bmatrix}
        \gls*{V_S}(k) \\
        \gls*{Q_S}(k) 
    \end{bmatrix} + 
    \begin{bmatrix}
        -\gls*{Q_S}(k) \gls*{Delta_t} \\
        \mathrm{ANN}_\mathrm{D}(\gls*{Q_S}(k), \gls*{u_S}(k), \gls{w_D})
    \end{bmatrix}
\end{multline}
contain the neural network $\mathrm{ANN}_\mathrm{D}$ with trainable parameters~\gls{w_D}.
The antechamber volume~$\gls{V_S}$ for the next time step~$k+1$ results from numerically integrating~$\gls{Q_S}(k)$ using the time step size~\gls{Delta_t}.

\subsection{Pressure Model}
\begin{figure}[ht]
    \centering
    \includegraphics[width=\columnwidth]{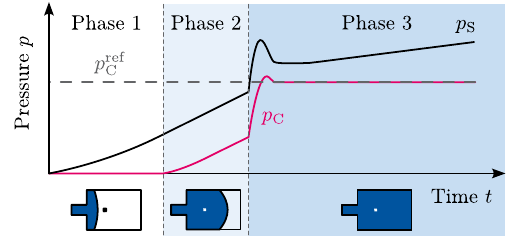}
    \caption{Schematic pressure curves of the cavity-pressure controlled IMP divided into three different phases depending on the degree of cavity filling (lower part of the figure).}
    \label{fig: DifferentPhasesOfIM}
\end{figure}

The process is divided into three phases, as shown in Fig.~\ref{fig: DifferentPhasesOfIM}, depending on the cavity fill level.
During injection, the mold fills with melt as the displaced screw volume~\gls{Delta_V_S} increases, with ~\mbox{$\gls{Delta_V_S}(t)~=~\int_{0}^{t}\gls{Q_S}(\tau)\,\mathrm{d}\tau$}.
% FIGURE: Motivation with different phases
Before the melt front reaches the cavity-pressure sensor, the sensor will measure ambient pressure~(Phase 1).
Once the melt front reaches the cavity-pressure sensor, the cavity pressure will increase moderately~(Phase 2). 
Once the mold is completely filled, there is no room for further melt expansion, so pressure increases rapidly~(Phase 3). 

\begin{figure}[ht]
    \centering
    \includegraphics[width=\columnwidth]{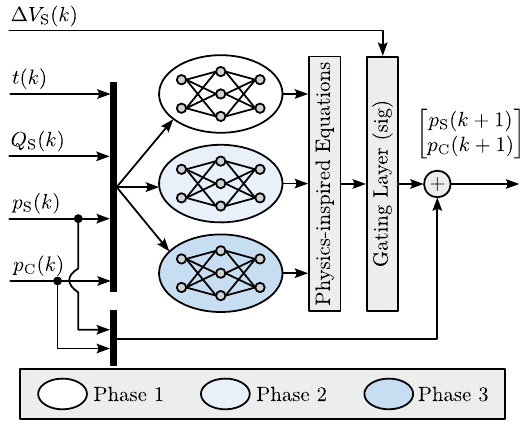}
    \caption{Approach for modeling the pressure dynamics of the IMP using a Physics-Inspired Neural Mixture-of-Local-Experts model.}
    \label{fig: PressureModel}
\end{figure}

The model depicted in Fig.~\ref{fig: PressureModel} aims to capture the three phases via three separate Local-Expert submodels.
These submodels smoothly transition via a sigmoid function $\mathrm{sig}(\cdot)$ as~\gls{Delta_V_S} increases, driven by the gating layer. 
The model outputs the pressures~$\gls{p_S}(k+1)$ and~$\gls{p_C}(k+1)$ for the next time step using a ResNet structure.
% FIGURE: Modeling
% Moved the fact that the temperature measurements are not available to this place: 
The model varies with time~$\gls{t}$ to approximate material cooling, as temperature measurements inside the mold are typically unavailable and provide limited insight into the material's core temperature.

To incorporate physics information, the neural network submodels do not calculate the residual of the pressures directly, but return weights~$\begin{bmatrix} a_1^{(i)} & a_2^{(i)} & a_3^{(i)} & a_4^{(i)} \end{bmatrix}$
that are used inside the equations
\begin{align}
    &\Delta \gls*{p_S}^{(i)}(k) = \frac{1}{\gls*{V_S}(k)}\Big(a_1^{(i)}(k) \gls*{Q_S}(k) - a_2^{(i)}(k)\Delta p(k) \Big) \\
    &\Delta \gls*{p_C}^{(i)}(k) = \begin{cases}
			  0, & i = 1\\
            a_3^{(i)}(k) \Delta p(k)  - a_4^{(i)}(k) , & i\in\{2,3\}\\
		 \end{cases}
    \label{eq: physics-constrainedEQ2}
\end{align}
to prescribe a physics-conform behavior for each submodel~\mbox{$i\in\{1,2,3\}$}.
The structure of these equations is inspired by well-known pressure-build-up equations also used in~\cite{vukovicAdaptiveModelbasedPredictive2022}. 
Using these equations and constraining the weights to~\mbox{$a^{(i)}(k)~>~0~~\forall{i,k}$}, using
\begin{equation}
    \text{softplus}(\cdot) = \log ( 1 + \exp(\cdot))   
\end{equation}
as the output-activation function, ensures physics-conform pressure-build-up.

The Mixture-of-Local-Experts model predicts pressure build-up as
% equation for the model update:
% Ist multline hier eher keine gute Idee? 
\begin{align}
    \begin{bmatrix}
        \gls{p_S}(k+1) \\
        \gls{p_C}(k+1) 
    \end{bmatrix} 
    & = \gls{f_P}(\gls{p_S}(k), \gls{p_C}(k), \gls{Delta_V_S}(k), \gls{Q_S}(k), \gls{t}(k)) \nonumber
    \\
    &=
    \begin{bmatrix}
        \gls{p_S}(k) \\
        \gls{p_C}(k)
    \end{bmatrix}  +
    \sum_{i=1}^{3} \psi_i(\gls{Delta_V_S}(k)) \cdot \begin{bmatrix}
        \Delta \gls{p_S}^{(i)} (k) \\
        \Delta \gls{p_C}^{(i)} (k)
    \end{bmatrix}
\end{align}
with the weights 
\begin{equation}
    \begin{bmatrix}
        \psi_1 \\
        \psi_2 \\
        \psi_3 \\
    \end{bmatrix}= \begin{bmatrix}
        1 - \mathrm{sig}(a [\gls{Delta_V_S} - s_1 ])  \\
        \mathrm{sig}(a [\gls{Delta_V_S} - s_1]) - \mathrm{sig}(a [\gls{Delta_V_S} - s_2])  \\
        \mathrm{sig}(a [\gls{Delta_V_S} - s_2]) \\
    \end{bmatrix}
\end{equation}
assigned to each individual local expert model at time step $k$  with the switching values \mbox{$s_1 < s_2$} given by the gating layer in Fig.~\ref{fig: PressureModel}.
The steepness factor $a$ scales the argument of the sigmoid function.
The architecture of the ANNs is defined by the hidden layer widths $L_1, L_2$, and $L_3$.

\subsection{Model Training} \label{sec: Model training}

We use the Levenberg-Marquardt (LM) algorithm to train the models with \gls{N_T} data points.
For the drive model, the one-step-prediction residual
\begin{equation}
    % Residual for fitting the drive model: 
     \gls{r_D}(k) =
            \begin{bmatrix}
        \gls{V_S}(k+1) \\
        \gls{Q_S}(k+1)
    \end{bmatrix} - \gls{f_D} \in \mathbb{R}^{2\times1}
\end{equation}
is inserted into the complete residual vector 
\begin{equation}
    \gls{R_D} = \begin{bmatrix}
        \gls{r_D}(1)^\top &
        \hdots &
        \gls{r_D}(\gls{N_T})^\top
    \end{bmatrix}^\top
    \in \mathbb{R}^{(2\cdot \gls{N_T} ) \times 1}.
\end{equation}
To prevent overfitting on the time variable, we add a term to the one-step residual of the pressure model
\begin{equation}
    \gls{r_P}(k) =
    \begin{bmatrix}

            \begin{bmatrix}
                \gls{p_S}(k+1) \\
                \gls{p_C}(k+1)
            \end{bmatrix} - \gls{f_P} \\
        \sqrt{\gls{lambda_Reg}} \cdot \frac{\partial \gls{f_P}}{\partial \gls{t}}
    \end{bmatrix}
    \in \mathbb{R}^{4\times1} ,
\end{equation}
regularizing high sensitivity of the ANN's prediction with respect to time by using the weight~\gls{lambda_Reg}.

The aggregated residual vector for the pressure model is written as.
\begin{equation}
    \gls{R_P} = \begin{bmatrix}
        \gls{r_P}(1)^\top &
        \hdots &
        \gls{r_P}(\gls{N_T})^\top
    \end{bmatrix}^\top
    \in \mathbb{R}^{(4\cdot \gls{N_T} ) \times 1}.
\end{equation}
All residual vectors, as well as the Jacobian matrices~$\nabla \gls{R_D}$ and~$\nabla \gls{R_P}$ with respect to the neural network parameters, are then used inside the LM algorithm to train the model.
We train the drive and pressure models after each cycle using the previous \gls{n_C} IM cycles.

\section{Controller Optimization} \label{sec: TrustRegionOptimization}

% controller -----------------------------------
\subsection{Generic Control Law}

% the control law:
The control law for a simple feedback controller can be written as
\begin{equation}
    \gls{u_S}(k) = \pi(\gls{x}(k), \boldsymbol{c}(k), \gls{p_C_ref}(k), \gls{Theta_C}) 
\end{equation}
with the vector of controller parameters \mbox{$\boldsymbol{\theta}_\mathrm{C} \in\mathbb{R}^{N_{\boldsymbol{\theta}_\mathrm{C}}}$}, which we aim to optimize.
The controller can also have internal states~$\boldsymbol{c}(k)$.
An update equation for these internal states can be written as
\begin{equation}
    % the internal controller states update equation: 
    \boldsymbol{c}(k+1) = \boldsymbol{\tau}(\boldsymbol{c}(k), \gls{x}(k), \gls{p_C_ref}(k), \gls{Theta_C}).
\end{equation}

% Residual Learning ---------------------------------

\subsection{Composite Cost Function}

To measure the performance of the controller over one cycle of length \gls{N_C}, we define the cost function
\begin{equation}
    J = \sum_{k=2}^{\gls{N_C}} \bigl(\gls{p_C}(k) - \gls{p_C_ref}(k) \bigr)^2 + R \bigl( \gls{u_S}(k) - \gls{u_S}(k-1) \bigr)^2,
    \label{eq: Cost Function}
\end{equation}
which penalizes tracking error and controller output deviation.
The controller output deviation is penalized with the weight \gls{R} to reduce strain on the injection molding machine's drive.

Combining the plant model and the controller allows predicting the costs $J_\mathrm{M}$, namely by inserting the simulated cavity-pressure curves and simulated controller outputs into~\eqref{eq: Cost Function}.
Since we do not assume that this simulated cost model perfectly matches the costs calculated from real plant experiments, we add a GP regression model to correct for this mismatch using previous observations.
The sum of the deterministic model $J_\mathrm{M}$ and the GP
% equation for the complete cost model: 
\begin{align}
&J_{\text{total}}(\gls{Theta_C})
=
J_M(\gls{Theta_C}) + \tilde J (\gls{Theta_C})
\\
&\tilde  J(\gls{Theta_C})\sim\mathcal{GP}\left(0,\,\gls{k_SE}(\gls{Theta_C},\gls{Theta_C}^\prime )\right)
\end{align}
is again a GP, and its predictive Gaussian distribution has mean~$\mathbb{E}(J_{\text{total}}(\gls{Theta_C}))$ and variance~$\mathrm{Var}(J_{\text{total}}(\gls{Theta_C}))$.
We denote the Squared Exponential (SE) kernel function with~$k_\mathrm{SE}$.

\subsection{Model-Guided Local Bayesian Optimization}
The MGLBO algorithm is provided as pseudocode in Alg.~\ref{alg:local_bo}.
Since we assume $J_\mathrm{M}$ and thus $J_\mathrm{total}$ is only accurate in the neighborhood of the last cycles, we use a trust-region-based approach.
We aim to find a controller parametrization yielding lower costs within the trust region around the current best parameter vector \gls{Theta_C_star}.
To obtain this, we use the TurBO-1 algorithm~\cite{erikssonScalableGlobalOptimization2019}.

% Trust region: 
We define the trust region at optimization iteration $n$ to be
\begin{equation}
    S = \left\{ \gls{Theta_C} \in \mathbb{R}^{N_{\boldsymbol{\theta}_\mathrm{C}}} \;| \; \gls{Theta_C_star} - \gls{b} \le \gls{Theta_C} \le \gls{Theta_C_star} + \gls{b} \right\},
\end{equation}
with the vector \gls{b} defining the trust box around \gls{Theta_C_star}, where the inequality holds element-wise.
If the algorithm determines controller parameters that yield lower costs, we increase the trust-region size and move its center to the new best controller parameters.
If it fails, we expect the model to be an inaccurate approximation to the real cost function and shrink $S$.
Therefore, the algorithm focuses on finding local optimal solutions. 

% Acquisition function: 
We determine the next evaluation point by minimizing the acquisition function
\begin{equation}
    \gls{alpha}(\gls{Theta_C}) = \mathbb{E}(J_\text{total}(\gls{Theta_C})) + \kappa  \sqrt{\text{Var}(J_\text{total}(\gls{Theta_C}))}~.
\end{equation}
This choice of the acquisition function trades off predicted performance and uncertainty, thereby enabling risk-aware controller tuning.
Higher values for $\kappa\in~\mathbb{R}^+$ result in a more cautious optimization. 
The optimizer
\begin{equation}
\gls{Theta_C_cand} = \mathop{\mathrm{argmin}}_{\gls{Theta_C} ~\in~ S} \left( \gls{alpha} (\gls{Theta_C}) \right) 
\end{equation}
is evaluated in the next IM cycle.

\begin{algorithm}
    \caption{Model-Guided Local Bayesian Optimization}
    \label{alg:local_bo}
    \begin{algorithmic}[1]
        \Require Current parameters \gls{Theta_C_cand}, Measured trajectory data $\mathcal{T} = \{\mathbf{X}, \boldsymbol{u}, \boldsymbol{c}\}$, Reference \gls{p_C_ref_vec}
        \Require Data History: $\boldsymbol{\theta}_{C, \mathrm{Data}}$, $J_{\mathrm{Data}}$, $\mathcal{T}_{\mathrm{Data}}$
        \Ensure Next parameter candidate $\gls{Theta_C}^{\mathrm{cand}}$

        \Statex

        \Procedure{Optimization Step}{$\gls{Theta_C}, \mathcal{T}, \gls{p_C_ref_vec}$}
        
            \State $J \gets \Call{calculate Costs}{\mathcal{T}, \gls{p_C_ref_vec}}$ \Comment{\eqref{eq: Cost Function}}

            \State Update Storage:
            \State $\boldsymbol{\theta}_{\mathrm{C}, \mathrm{Data}} \gets \boldsymbol{\theta}_{\mathrm{C}, \mathrm{Data}} \cup \{\gls{Theta_C}\}$ \Comment{Store parameters}
            \State $J_{\mathrm{Data}} \gets J_{\mathrm{Data}} \cup \{J\}$ \Comment{Store costs}
            \State $\mathcal{T}_{\mathrm{Data}} \gets \mathcal{T}_{\mathrm{Data}} \cup \{\mathcal{T}\}$ \Comment{Store trajectory}
            \Statex

            \State Extract training set $\mathcal{T}_{\mathrm{train}} \subset \mathcal{T}_{\mathrm{Data}}$ (last $\gls{n_C}$ cycles)
            \State Fit Model $\mathcal{M}$ on $\mathcal{T}_{\mathrm{train}}$ \Comment{Sec. \ref{sec: Model training}}

            \Statex

            \If{$J < J^* - \epsilon$} \Comment{Improvement}
                \State $\gls{Theta_C_star} \gets \gls{Theta_C_cand}$; $J^* \gets J$; $\mathcal{T}^* \gets \mathcal{T}$
                \State $\gls{b} \gets \gls{b} \cdot \gls{kappa_suc}$ \Comment{Expand Trust Region}
            \ElsIf{$J \leq J^* + \epsilon$} \Comment{Neutral / Stagnation}
                \State $\gls{Theta_C_star} \gets \gls{Theta_C_cand}$; $J^* \gets \min(J, J^*)$; $\mathcal{T}^* \gets \mathcal{T}$
            \Else \Comment{Performance worse}
                \State $\gls{b} \gets \gls{b} \cdot \gls{kappa_fail}$ \Comment{Shrink Trust Region}
            \EndIf

            \Statex

            \State Define function $J_M(\gls{Theta_C}) := \Call{Simulate}{\mathcal{M}, \gls{Theta_C}}$

            \Statex

            \State Fit $\Tilde{J}$ to residuals $(J_{\mathrm{Data}} - \gls{J_M}(\boldsymbol{\theta}_{C, \mathrm{Data}}))$
            \State $\gls{Theta_C_cand} \gets \text{argmin}_{\gls{Theta_C}} \gls{alpha}(\gls{Theta_C})$ s.t. $\gls{Theta_C} \in S$
            \State \Return \gls{Theta_C_cand}
            
        \EndProcedure 
        
    \end{algorithmic}
\end{algorithm}

\section{Simulation Setup} \label{sec:simulated-experiments}

% Introducing the evaluation model and naming differences: 
The first-principles plant model used for evaluation is described in~\cite{ahlersPartMassControlInjection}.
Unlike the model presented in Sec.~\ref{sec:process-modeling}, it is not separated into distinct phases and relies on first-principles modeling.
It accounts for non-Newtonian fluid behavior, assuming a power-law fluid, time-dependent cooling of the polymer melt, and a mold-specific bulk modulus that varies with the displaced screw volume.

% Proportional Controller: 
The control law of a P controller is defined by
\begin{equation}
    \pi_\mathrm{P}(k) = [\gls{Theta_C}]_1 \cdot (\gls{p_C_ref}(k)-\gls{p_C}(k)),
\end{equation}
which only includes one tunable parameter.

% Gain-Scheduled Controller: 
The control law and the internal-state-update equation for the G-PI controller are written as
\begin{align}
    &\pi_\mathrm{PI}(k) = \gls{K_P}^{\mathrm{eff}} e(k) + K_\mathrm{I}^{\mathrm{eff}} c_\mathrm{PI}(k) \\
    &c_\mathrm{PI}(k+1) = c_\mathrm{PI}(k) + \Delta t \cdot [\boldsymbol{w}_\mathrm{P}]_3 \cdot e(k) \\
    &e(k) = \gls{p_C_ref}(k)-\gls{p_C}(k)
\end{align}
with the effective proportional gain $\gls{K_P}^\mathrm{eff}$ and the effective integral gain $K_\mathrm{I}^{\mathrm{eff}}$.
The weights 
\begin{equation}
    \boldsymbol{w}_\mathrm{P} = \begin{bmatrix}
        1 - \mathrm{sig}(a (\gls{Delta_V_S} - [\gls{Theta_C}]_5))  \\
        \mathrm{sig}(a (\gls{Delta_V_S} - [\gls{Theta_C}]_5)) - \mathrm{sig}(a (\gls{Delta_V_S} -[\gls{Theta_C}]_6))  \\
        \mathrm{sig}(a (\gls{Delta_V_S} - [\gls{Theta_C}]_6)) \\
    \end{bmatrix}
\end{equation}
are used to determine the effective gains, with a factor for the steepness $a$.
The effective proportional and integral gain of the controller
\begin{align}
    &\gls{K_P}^{\mathrm{eff}} = \boldsymbol{w}_\mathrm{P}^\top \cdot 
    [\gls{Theta_C}]_{1:3} \\
    &K_\mathrm{I}^{\mathrm{eff}} = [\boldsymbol{w}_\mathrm{P}]_3 \cdot [\gls{Theta_C}]_4
\end{align}
depend on the displaced screw volume \gls{Delta_V_S}.

% Radial basis function controller: 
The RBF controller exemplifies a highly nonlinear control law with more tunable parameters than the aforementioned controllers. 
The control law
\begin{align}
    &\pi_\mathrm{RC}(k)  = \gls{Theta_C}^\top \boldsymbol{\psi} \\
    &\psi_i(k) = \exp\Biggl(- \frac{1}{2 \sigma_{\mathrm{RC}}^2}\biggl(\frac{\gls{p_C}(k)}{\gls{p_C_ref}(k)} - [\boldsymbol{r}_\mathrm{RC}]_i\biggr)^2 \Biggr)
\end{align}
includes $N_\mathrm{RC}$ basis functions located at the center points~$[\boldsymbol{r}_\mathrm{RC}]_i$ with a measure for their width~$\sigma_\mathrm{RC}$.

% explain how the controllers are initialized: 
The controllers are parameterized such that the initial controller is always a P controller, with gain $K_\mathrm{P}^{\mathrm{Init}}$. 
The idea is that a P controller for the IMP is easy and safe to tune by incrementally increasing its proportional gain, and therefore serves as the starting point for each optimization.
For the G-PI controller, enforcing the P control law is straightforward.
For the RBF controller, we fitted the weights using least squares and the pseudo-inverse to the P control law.

% Limitations of the drive: 
Given that a negative~\gls{u_S} drives a positive~\gls{Q_S}, the control signal is constrained to the interval $\gls{u_S} \in [\gls{u_S_lowerBound}, \gls{u_S_upperBound}]$. 
The lower bound addresses the maximum injection speed of the machine, while the upper bound prevents undesirably high screw retraction.

% Baseline optimization algorithms: 
To compare MGLBO against a baseline controller-tuning method, we use the MATLAB-internal \textit{bayesopt} function with the \textit{expected-improvement-plus} acquisition function~\cite{StatisticsMachineLearning2024}.
For the Vanilla BO, we define bounds $\gls{Theta_C}^\pm$ for each controller to limit the search space.

% Cost normalization: 
We normalize all costs by the costs achieved by using the P controller with the initial gain $K_\mathrm{P}^{\mathrm{Init}}$.

% Performance metrics: 
To assess MGLBO's performance, we compare it using three metrics. 
\begin{enumerate}
    \item We evaluate the controller performance using the minimum observed cost~$J_\mathrm{min}$. This comparison is particularly useful for determining whether the local exploitation of MGLBO can achieve solution quality comparable to that of global optimization in Vanilla BO.
    \item The maximum cost~$J_\mathrm{max}$ is used to evaluate the safety of the tuning process. Monitoring this metric is critical to industrial deployment, as unsafe controller parameters risk causing physical damage to the machine.
    \item The metric of cumulative worsening 
    \begin{equation}
        J_\mathrm{W}(n) = \sum_{i = 2}^{n} \max (0, J_i - J_{i-1})
    \end{equation}
    is also important in industrial applications. 
    If machine operators observe frequent deterioration in controller performance, they may disable the algorithm to protect the IM machine.
\end{enumerate}

In Tab.~\ref{tab:DetailedParameters}, we visualize the parameters used for the investigation.

% Table with the used parameters: 
\begin{table}[hbt!]
    \centering
    \caption{Parameters used for the investigation.}
    \label{tab:DetailedParameters}
    
    % --- HORIZONTAL TIGHTENING ---
    % Reduces the whitespace between columns. 
    % Default is usually 6pt. We set it to 1.5pt.
    \setlength{\tabcolsep}{2pt} 
    \renewcommand{\arraystretch}{1.05}
    
    % l = left (Symbol), l = left (Name), c = center (Value), c = center (Unit)
    \begin{tabular}{llcc}
        \toprule
        \textbf{Symbol} & \textbf{Parameter Name} & \textbf{Value} & \textbf{Unit} \\
        \midrule

         % --- SECTION ---
        \multicolumn{4}{l}{\textit{\textbf{Mixture-of-Local-Experts Model}}} \\
        \addlinespace[0.5ex]
        
        $\gls{n_LM}$           & Number of LM steps                   & 50        & -  \\
        $\gls{n_C}$            & Number of training cycles  & 4         & -  \\
        $L_{\mathrm{D}}$       & Hidden layers $\mathrm{ANN}_{\mathrm{D}}$           & [5]       & -  \\
        $L_1$                  & Hidden layers $\mathrm{ANN}_{\mathrm{1}}$           & [5]       & -  \\
        $L_2$                  & Hidden layers $\mathrm{ANN}_{\mathrm{2}}$            & [5]       & -  \\
        $L_3$                  & Hidden layers $\mathrm{ANN}_{\mathrm{3}}$           & [5~ 5]    & -  \\

        \addlinespace[1.5ex]

        % --- SECTION ---
        \multicolumn{4}{l}{\textit{\textbf{MGLBO}}} \\
        \addlinespace[0.5ex]
        
        $\kappa_\mathrm{suc}$      & Increasing factor of \gls{b}       & $1.1$            & -          \\
        $\kappa_\mathrm{fail}$     & Decreasing factor of \gls{b}        & $0.9$            & -          \\
        $\epsilon$                 & Cost tolerance                     & $1\cdot 10^{-5}$ & -          \\
        $R$                        & Penalty for input deviation        & $5\cdot 10^{6}$  & \si{V^{-2}} \\ 
        $\kappa$                   & Penalty for uncertainty            & $1.645$          & - \\ 
        
        \addlinespace[1.5ex]
        
        % --- SECTION ---
        \multicolumn{4}{l}{\textit{\textbf{P controller}}} \\
        \addlinespace[0.5ex]
        $\gls{Theta_C}^\pm$        & Bounds for \gls{Theta_C}             & $[0.1 ~~ 1] \cdot10^{-2}$  & \si{\V\bar^{-1}} \\  
        \addlinespace[0.25ex]
        $\gls{b}^{0}$  & Initial trust-region size            & $1\cdot10^{-3}$  & \si{\V\bar^{-1}} \\ 
    
        \addlinespace[1.5ex]
    
        % --- SECTION ---
        \multicolumn{4}{l}{\textit{\textbf{G-PI controller}}} \\
        \addlinespace[0.5ex]

        $[\gls{Theta_C}^\pm]_{1:3}$              & Bounds for $[\gls{Theta_C}]_{1:3}$         & $[0.1 ~~ 1] \cdot10^{-2}$  & \si{\V\bar^{-1}}   \\
        \addlinespace[0.25ex]
        $[\gls{Theta_C}^\pm]_4$                & Bounds for $[\gls{Theta_C}]_{4}$            & $[0~~ 5] \cdot10^{-3}$    & \si{\V\bar^{-1}\s^{-1}}        \\  
        \addlinespace[0.25ex]
        $[\gls{Theta_C}^\pm]_5$                & Bounds for $[\gls{Theta_C}]_{5}$           & $[25~~ 35]$               & \si{\cm^3}         \\ 
        \addlinespace[0.25ex]
        $[\gls{Theta_C}^\pm]_6$                & Bounds for $[\gls{Theta_C}]_{6}$           & $[45~~ 55]$               & \si{\cm^3}         \\  
        \addlinespace[0.25ex]
        $[\gls{b}^{0}]_{1:3}$                    & Initial trust-region size          & $1\cdot10^{-3}$  & \si{\V\bar^{-1}} \\ 
        $[\gls{b}^{0}]_{4}$                      & Initial trust-region size          & $1\cdot10^{-4}$  & \si{\V\bar^{-1}\s^{-1}} \\ 
        $[\gls{b}^{0}]_{5:6}$                    & Initial trust-region size          & $2$              & \si{\cm^3}  \\ 

        \addlinespace[1.5ex]

         % --- SECTION ---
        \multicolumn{4}{l}{\textit{\textbf{RBF controller}}} \\
        \addlinespace[0.5ex]

        $N_\mathrm{RC}$                   & Number of RBF                       & $30$           & -               \\
        $\sigma_\mathrm{RC}$              & RBF Standard deviation              & $0.05$         & -               \\ 
        $\gls{Theta_C}^\pm$               & Bounds for \gls{Theta_C}       & $[-1.5~~ 0.5]$  & \si{\V}         \\
        $\gls{b}^{0}$                     & Initial trust-region size           & $0.1$          & \si{\V}         \\
        
        \addlinespace[1.5ex]
        
        % --- SECTION ---
        \multicolumn{4}{l}{\textit{\textbf{General}}} \\
        \addlinespace[0.5ex]

        $\Delta t$      & Time step size             & $8\cdot10^{-3}$    &  \si{s}  \\
        \gls{p_C_ref}   & Cavity-Pressure reference  & $250$  &  \si{bar} \\
        $\gls{N_C}$     & Cycle length               & $1000$ &  -         \\
        $s_{1}$                               & First switching point               & $30$              & \si{\cm^3}  \\
        $s_{2}$                               & Second switching point              & $50$              & \si{\cm^3}  \\
        $a$                               & Steepness for sig function              & $4$               & \si{\cm^{-3}}    \\
        $K_\mathrm{P}^{\mathrm{Init}}$    & Initial proportional gain               & $4\cdot10^{-3}$    & \si{\V  \bar^{-1}}  \\

        \bottomrule
    \end{tabular}
\end{table}

\section{Results} \label{sec:results}

\begin{figure}[ht]
     \centering
     \includegraphics[width=\columnwidth]{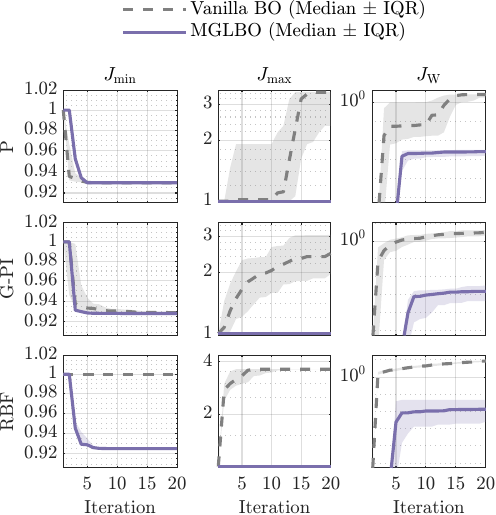}
     \caption{Comparison between the MGLBO and Vanilla BO for~$J_\mathrm{min}$,~$J_\mathrm{max}$,~and~$J_\mathrm{W}$ for 50 individual runs.}
     \label{fig:ComparisonErrorMetrics}
\end{figure}

% Results of the optimization: 
Fig.~\ref{fig:ComparisonErrorMetrics} illustrates the distributions for the metrics from Sec.~\ref{sec:simulated-experiments} for both MGLBO and Vanilla BO, derived from 50 independent runs per controller. 
In the initial iteration, all controllers are set to the P control law.
From the plot of~$J_\mathrm{min}$ in Fig.~\ref{fig:ComparisonErrorMetrics}, it can be seen that the controllers with higher parameter counts yield lower overall minimum costs.

% Figure comparing the MGLBO with the Vanilla BO: 

% Talking about the MGLBO: 
For the MGLBO, all second controller generations yield only minor improvements over the first ones. 
During the MGLBO, all controllers converge to a final parametrization.
The minimum costs obtained by MGLBO are lower than or equal to those of Vanilla BO, while the maximum costs correspond to the initial costs.
The first deterioration of the controllers begins at iteration five, coinciding with the algorithm approaching convergence.

% Talking about the Vanilla BO: 
For optimization using Vanilla BO, we observe that for the P controller, Vanilla BO shows an initial improvement higher than that of MGLBO. 
For all other controllers, the optimization converges more slowly compared to MGLBO. 
For the RBF controller, the Vanilla BO does not identify a better parameterization than the initial one within 20 iterations.
Overall, the cumulative worsening of the controller's performance for Vanilla BO is substantially greater than for MGLBO.

\begin{figure}[ht!]
     \centering
     \includegraphics[width=\columnwidth]{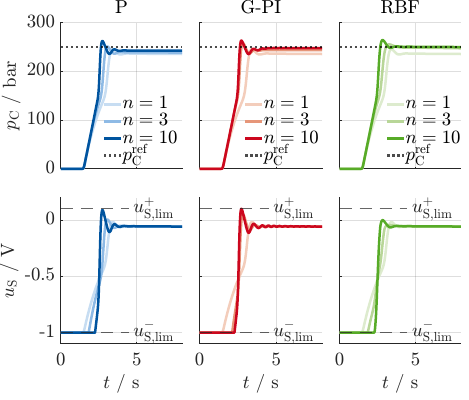}
     \caption{Comparison between the cavity-pressure curves and control inputs resulting from different optimization iterations $n$ for all three controllers.}
     \label{fig:pCCurves}
\end{figure}

% Showing the figure containing the pressure curves: 
In Fig.~\ref{fig:pCCurves}, we visualize the cavity-pressure curves and the control inputs during the optimization process with MGLBO for three different iterations. 
% Hier erkenne ich mit Coblis deutliche Unterschiede durch die Intensität.
All controllers result in the same cavity-pressure curves at the first iteration of the optimization, as they are initialized using the P control law.

Comparing the converged cavity-pressure curves of the P controller and the G-PI controller, we observe that the G-PI controller does not exhibit a stationary offset.
The RBF controller also does not exhibit a stationary offset and exhibits a reduced cavity-pressure overshoot.
Additionally, it can be seen that the RBF controller does not reach the upper bound \gls{u_S_upperBound} of the controller output.

\begin{figure}[ht]
    \centering
    \includegraphics[width=\columnwidth]{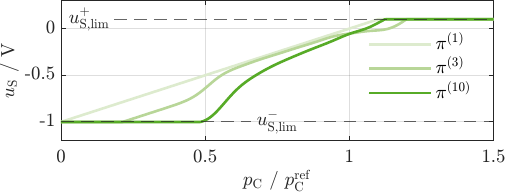}
    \caption{The control law $\pi^{(n)}$ for the RBF controller for three different iterations as a function of the nondimensional cavity pressure starting as a P controller.}
     \label{fig:RBFControllerCurve}
\end{figure}

In Fig.~\ref{fig:RBFControllerCurve}, we visualize the control law obtained by optimizing the RBF controller. 
The control law again starts with a P controller, successively adding nonlinearity to $\pi$. 
% Hier erkenne ich mit Coblis deutliche Unterschiede durch die Intensität
For lower cavity pressures, the RBF controller increases the screw velocity by reducing \gls{u_S} compared to the initial control law.
For cavity-pressure values close to the cavity-pressure reference, the control law of the RBF controller approaches the initial proportional control law.
The control law of the RBF controller exhibits a negative offset for $\gls{p_C} = \gls{p_C_ref}$, unlike the P control law, where the P controller output is exactly zero. 

% Computation time
During the optimization for 20 MGLBO steps, the mean execution time is \SI{9.4}{\s} on an 11th Gen Intel(R) Core(TM) i7-1165G7 with \SI{2.8}{\GHz} running in MATLAB R2024b~\cite{MATLAB2024} for the optimization of the RBF controller.

\section{Discussion} \label{sec:discussion}

\subsection{Potential for Controller Optimization}

The MGLBO approach demonstrates strong potential for risk-aware, data-efficient controller optimization in the IMP.
Adding model information enables a more targeted optimization than Vanilla BO, with lower maximum costs and lower cumulative worsening.
For a higher number of parameters, such as the RBF controller, the advantages of the composite objective function also become apparent for $J_\mathrm{min}$, since the prediction of the costs using the dynamic model is not influenced by a higher parameter count~$N_{\boldsymbol{\theta}_\mathrm{C}}$. 

% Small steps at the beginning: 
The algorithm takes very small steps at the beginning of the optimization because the value-at-risk acquisition function is employed, and the GP model relies on the initial observation.
This behavior could be improved by conducting additional prior experiments to obtain an initial set of observations, or by making a reasonable guess about the GP hyperparameters.

% Global vs. Local: 
The Vanilla BO does not find lower minimum costs for P and G-PI.
This may indicate that global optimization methods are not required to identify optimal, interpretable cavity-pressure controllers.
Global optimization methods search the entire domain of admissible solutions. 
Therefore, they may be unsuitable for learning the control law for cavity-pressure controllers, as poor parameterization can damage the IM machine and lead to scrap parts. 

% Justifying the increased computational effort:
Although the computational effort of MGLBO is substantially higher than that of Vanilla BO, particularly due to model retraining, the optimization is allowed to run for the full cooling time.
The cooling time usually makes up more than half of the complete cycle~\cite{hopmannTechnologieSpritzgiessensLern2017}.
For larger models that require more training time, updating model parameters over multiple cycles may also be sufficient.
The optimization time can also be reduced by using parallel computing and by increasing termination tolerances and reducing the number of iterations of the underlying optimizations.

% Differences to real world: 
The proposed approach must be adapted for the industrial application.
The transition points between the local expert models are not updated in this work and are assumed to be known.
For a real-world application, these points must be updated, by e.g., using the Maximum-Expectation Algorithm~\cite{dempsterMaximumLikelihoodIncomplete1977} or an Interacting Multiple Model filter~\cite{blomInteractingMultipleModel1988}.
% Number of states
Although the drive model can capture nonlinear dynamics, in its current form, it is limited to single-state drives. 
While this may be sufficient for IMs with electric drives, modeling IMs with hydraulic drives may require adding additional states~\cite{froehlichControlorientedModelingServopump2018}.

\subsection{Outlook}

% Testing on real hardware: 
Future work will involve testing the algorithm on real IM hardware under realistic conditions and comparing different interpretable control laws with model-based controllers.

% Incorporating other things into the algorithm
Beyond guiding controller optimization, incorporating a process model into the optimization loop enables additional options. 
One could add process-dependent constraints, such as maximum admissible cavity-pressure overshoots.
Combined with quality-attribute models that use the plant's simulated pressure trajectories, we could also treat quality-attribute references as equality constraints in the controller optimization.
This could yield a tightly controlled IMP while meeting desired quality-attribute references.

\addtolength{\textheight}{-12cm}   % This command serves to balance the column lengths

\bibliographystyle{IEEEtran}
\bibliography{combined-optimizer}

\end{document}